\documentclass[useAMS,usenatbib]{mn2e}
\usepackage{aas_macros}
\usepackage{graphicx}
\usepackage{amssymb}
\usepackage{amsmath}
\usepackage{widetext}
\def\psad{$P^2SAD$}
\usepackage{color}

\title[Clustering in the Phase Space of Dark Matter Haloes. II. Stable Clustering and Dark Matter Annihilation]
{Clustering in the Phase Space of Dark Matter Haloes. II. Stable Clustering and Dark Matter Annihilation}
\author[Jes\'us Zavala, Niayesh Afshordi]{\parbox{18cm}{Jes\'us
    Zavala$^{1,2,3,4}$\thanks{e-mail: jzavala@dark-cosmology.dk} and Niayesh Afshordi$^{1,2}$\vspace{0.3cm}}\\ 
$^{1}$Perimeter Institute for Theoretical Physics, 31 Caroline St. N., Waterloo, ON, N2L 2Y5, Canada\\
$^{2}$Department of Physics and Astronomy, University of Waterloo, Waterloo, Ontario, N2L 3G1, Canada\\
$^{3}$Canadian Institute for Theoretical Astrophysics, University of Toronto, Toronto, Ontario M5S 3H8, Canada\\
$^{4}$Dark Cosmology Centre, Niels Bohr Institute, University of Copenhagen, Juliane Maries Vej 30, 2100 Copenhagen, Denmark\\}

\begin{document}



\maketitle

\label{firstpage}

\begin{abstract}
We present a model for the structure of the 
particle phase space average density (\psad) in galactic haloes, 
introduced recently as a novel measure of the clustering of dark matter. Our model is based on the stable clustering hypothesis 
in phase space, the spherical collapse model, and tidal disruption of substructures, which is calibrated against the 
Aquarius simulations. Using this 
model, we can 
predict the behaviour of \psad~in the numerically unresolved 
regime, down to the decoupling mass limit of generic WIMP models. This prediction can be used to estimate signals sensitive to the
small scale structure of dark matter. 
For example, the dark matter annihilation rate 
can be estimated for arbitrary velocity-dependent cross sections in a convenient way
using a limit of \psad~to zero separation in physical space.
We illustrate our method by computing the 
global and local subhalo 
annihilation boost to that of the smooth dark matter distribution in a Milky-Way-size halo. Two cases are considered, one where
the cross section is velocity independent and one that approximates Sommerfeld-enhanced models. We find that the global
boost is $\sim10-30$, which is at the low end of current estimates (weakening expectations of large extragalactic signals), 
while the boost at the solar radius is below the percent level. We make our code to compute \psad~publicly available, which can be used
to estimate various observables that probe the {\it nanostructure} of dark matter haloes.  
\end{abstract}


\section{Introduction}

Despite being the most dominant type of matter in the Universe, dark matter is evident so far only through its gravitational effects on
luminous matter. The nature of dark matter will thus continue to be elusive unless it has detectable non-gravitational interactions. Weakly
Interacting Massive Particles (WIMPs) are among the favourite dark matter candidates and are expected to give promising signals 
that can be detected
experimentally either directly through scattering off nuclei in laboratory detectors, or indirectly through the byproducts of their
self-annihilation into standard model particles (e.g. photons and electron/positrons pairs). Several experiments 
are pursuing such a discovery and their sensitivity is reaching the level of interaction predicted by popular WIMP models
\citep[e.g.][]{Abazajian_12,Aprile_12}. 

Interestingly, there are tantalizing anomalies that might be caused by new dark matter physics.
For instance, although the excess of $>10$~GeV cosmic ray positrons established firmly by the PAMELA and Fermi Satellites 
\citep{Adriani_09,Abdo_09}, and recently confirmed with high precision by the AMS collaboration \citep{AMS_13}, 
can be explained by ``ordinary'' astrophysical sources (e.g. nearby pulsars, \citealt{Linden_13}), it can also
be interpreted as dark matter annihilation.
Large annihilation rates, of $\mathcal{O}(100-1000)$ above the expected rate of a $\sim$TeV mass thermal relic, and primarily leptonic 
final states are needed, however, to explain the data \citep{Bergstrom_09}. These requirements can be satisfied by leptophilic 
dark matter models coupled to light force carriers that enhance the annihilation cross section via a Sommerfeld mechanism 
\citep[e.g.][]{Arkani_09,Pospelov_09}. An anomalous extended gamma-ray emission at intermediate galactic latitudes (peaking at $1$~GeV) 
has also been interpreted as a signal of dark matter annihilation \citep[e.g.][]{Hooper_13}.

Predictions for the hypothetical non-gravitational signatures of dark matter are highly dependent on the clustering of dark matter at
small scales. Although the steady progress of numerical $N-$body simulations over the past few decades has given us a detailed picture
of the spatial dark matter distribution from large ($\sim$~Gpc) to sub-galactic scales ($\sim100$~pc), the regime most relevant for
certain dark matter detection efforts (those based on extragalactic signals) remains below the resolution of current simulations 
($\mathcal{O}(10^3$M$_\odot$) at $z=0$; \citealt{Springel_08,Diemand_08,Stadel_09}). This is because the current
cold dark matter (CDM) paradigm of structure formation predicts a hierarchical scenario with a very large mass range of gravitationally
bound dark matter structures, from $10^{15}$M$_\odot$ cluster-size haloes down to a decoupling mass limit of $10^{-11}-10^{-3}$M$_\odot$
\citep[e.g.][]{Bringmann_09}. The characteristic sizes of these small haloes (commonly called microhaloes) varies from $\sim10^{-7}-10^{-10}$ times the
size of a Milky Way (MW) type halo (we call this the {\it nanostructure} of dark matter haloes).

Despite their limited resolution, current simulations clearly suggest that the contribution of low-mass (sub)haloes is  
dominant for dark matter annihilation signals in the case of extragalactic sources, such as gamma-rays from 
galaxy clusters \citep[e.g.][]{Gao_12} and from integrated backgrounds \citep[e.g.][]{Zavala_10,Fornasa_13}. Within the solar radius, 
it seems that the dominant signal should come from the diffuse distribution of dark matter rather than by small scale subclumps 
\citep{Springel_08b}. Substructures within the MW halo can be detected in gamma-rays
either individually as the hosts of satellite galaxies \citep{Ackermann_11}, dark satellites devoid of stars \citep{Ackermann_12} or 
as a dominant contribution to the angular power spectrum of a full-sky gamma-ray signal \citep[e.g.][]{Siegal_08}.

Extrapolations below resolved scales are therefore needed to obtain a prediction of the expected signals from dark matter annihilation. 
For a given host halo, these extrapolations ultimately depend on the survivability of the smallest substructures as they are tidally
disrupted by the host. Low mass subhaloes form earlier in the hierarchical scenario and, being the densest, are expected to
survive tidal disruption contributing heavily to the annihilation signal. Current estimates on this contribution mostly rely on assumptions about the
abundance, spatial distribution, and internal properties of unresolved subhaloes that could lead to significant over/underestimations.
Typical estimates calibrated to simulation data vary by up to an order of magnitude \citep[e.g.][]{Springel_08,Kuhlen_08,Kamionkowski_10}.
The role of substructure, and therefore the uncertainty in the extrapolations to the
unresolved regime, is magnified in Sommerfeld-enhanced models where the annihilation 
cross section scales as the inverse velocity making the cold and dense subhaloes dominant not only in extragalactic structures 
\citep{Zavala_11} but plausibly also locally \citep{Slatyer_12}. 

In this paper, we present a novel method to study dark matter clustering that can be used to
estimate the dark matter annihilation rate calibrated with simulation data. This method is based 
on a physically-motivated model inspired by an extension into phase space of
the stable clustering hypothesis \citep{Afshordi_10}, originally introduced in position space by \citet{Davis_77}. 
The novelty of the method relies on using a more complete picture of dark matter clustering based on
a new quantity, the particle phase space average density (\psad), a coarse-grained phase space density that can be used straightforwardly 
to compute the annihilation rate. \psad~was introduced in a companion paper \citep[][hereafter Paper I]{Zavala_13} where we studied 
its main features in galactic-size
haloes using the simulation suite from the Aquarius project \citep{Springel_08}. Remarkably, we found it to be nearly universal at small scales
(i.e. small separations in phase space) across time and in regions of substantially different ambient densities.
This near universality is also preserved across the different Aquarius haloes, having similar masses but diverse mass accretion histories: 
they possess only slightly different \psad s at small scales without any re-scaling.
We argue in this paper that this behaviour can be roughly described by a refinement of the stable clustering hypothesis through
a simple model that incorporates tidal disruption of substructures.

In cases other than s-wave self-annihilation, the dependence of the interaction on the 
relative velocity of the annihilating particles has to be averaged over the velocity distribution of the dark matter particles. 
Although it is common to assume a Maxwellian velocity distribution, current simulations have shown that there are significant deviations
from this assumption related to the dark matter assembly history \citep[e.g.][]{Vogelsberger_09}. Our method is particularly useful 
for models where the cross section is velocity dependent (such as in Sommerfeld-enhanced models) because the particle phase space average 
density (\psad) that we introduce does not rely on assumptions about the velocity distribution and can deal with these cases in a natural way.

This paper is organised as follows: In Section \ref{2PCF}, we summarize the main results of Paper I, introducing the clustering
of dark matter through \psad~and its nearly universal structure at small scales in galactic haloes. In Section \ref{sec_rate}, we describe
how the annihilation rate is directly connected to the limit of zero physical separation of \psad. In Section \ref{stable_model}, we describe
our model of the small scale structure of \psad~based on the spherical collapse model and the stable clustering hypothesis in phase space,
refined by a subhalo tidal disruption prescription. In Section \ref{sim_fit}, we fit our model to the simulation data while in Section \ref{boost_sec}
we illustrate how this can be used to compute the global and local subhalo boost to the smooth annihilation rate in a MW-size halo. Finally, 
a summary and our main conclusions are given in Section \ref{conc_sec}.

\section{The Particle Phase Space Average Density (\psad) on small scales}\label{2PCF}

We follow the same definitions as in Paper I and study the clustering in phase space through 
\psad, defined as the mass-weighted average (over a volume ${\cal V}_6$ in phase space) 
of the coarse-grained phase space density of dark matter, on spheres of radius $\Delta x$ and $\Delta v$, in position and velocity 
spaces, respectively:
\begin{equation}\label{coarse}
\Xi(\Delta x, \Delta v)\equiv
\frac{\int_{{\cal V}_6}d^3{\bf x}d^3{\bf v} f({\bf x},{\bf v})f({\bf x}+{\bf\Delta x},{\bf v}+{\bf \Delta v})}
{\int_{{\cal V}_6} d^3{\bf x}d^3{\bf v} f({\bf x},{\bf v})}
\end{equation}
where $f({\bf x},{\bf v})$ is the phase space distribution function at the phase space coordinates $\bf x$ and $\bf v$.

In Paper I of this series we implemented and tested an estimator of \psad~in $N-$body simulations based on pair counts and applied to the
set of Aquarius haloes \citep{Springel_08}. In this paper, we refer exclusively to the results found for halo Aq-A-2 that has a virial mass and radius,
defined with a mean overdensity of 200 times the critical value, of $1.8\times10^{12}~$M$_\odot$ and $246$~kpc, respectively. 

One of the main results that we obtained is that the structure of \psad~averaged within the virialized halo is clearly separated by
two regimes: (i) a region at large scales (i.e. large separations in phase space) dominated by the smooth dark matter distribution where
\psad~varies in time due to the inside-out growth of the dark matter halo, and (ii) a region at small scales (i.e. small separations in 
phase space) dominated by gravitationally bound substructures where \psad~is nearly universal across time and ambient density. In the reminder
of this paper we consider only the small scale regime, which is the one of relevance in estimating the impact of substructure on
dark matter annihilation signals.

\begin{figure}
\center{
\includegraphics[height=8.0cm,width=8.0cm]{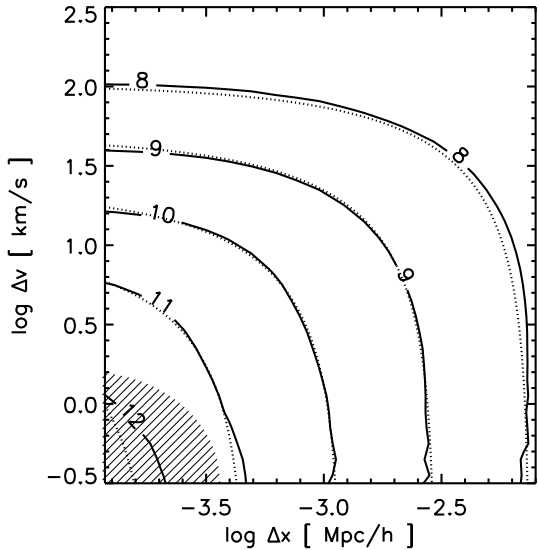} 
}
\caption{Contours of the logarithm of the particle phase space average density (\psad) of the Aq-A-2 halo (solid) at $z=0$ and for the fitting function
defined in Eqs.~(\ref{lame}-\ref{lame_2}) (dashed) with the parameters as given in Table \ref{table_fit_0}.
The dashed region in the left corner encompasses the region where resolution is a potential issue (see Appendix of Paper I).} 
\label{Fig_comp_fit} 
\end{figure}

Although the small scale regime can be roughly described by a subhalo model with abundance and properties given by scaling laws
fitted to the simulation data (see Fig. 5 of Paper I), we found that a better description is given by a family of superellipse contours 
of constant \psad$\equiv\Xi$, with parameters that can be adjusted to fit the variations of \psad~in redshift and halo-centric distance (see Tables
2 and 3 of Paper I):
\begin{equation}\label{lame}
\left(\frac{\Delta x}{\mathcal{X}(\Xi)}\right)^{\beta} + \left(\frac{\Delta v}{\mathcal{V}(\Xi)}\right)^{\beta} = 1,
\end{equation}
where $\beta(z;r)$ is a shape parameter and $\mathcal{X}(\Xi)$ and $\mathcal{V}(\Xi)$ are the axes of the superellipse: 
\begin{eqnarray}\label{lame_2}
\mathcal{X}(\Xi)=q_X(z;r)\Xi^{\alpha_X(z;r)}\\
\mathcal{V}(\Xi)=q_V(z;r)\Xi^{\alpha_V(z;r)}\nonumber
\end{eqnarray}
Fig.~\ref{Fig_comp_fit} shows contours of log(\psad) in the small scale regime averaged within the virialized region of Aq-A-2 at $z=0$ (solid lines). The
dotted lines are the fitting contours through Eqs.~(\ref{lame}-\ref{lame_2}) with the parameters as given in Table \ref{table_fit_0}. 
In section \ref{stable_model}, we present the model that motivates this fitting formula. We note that the shaded area on the left corner encompasses the
region where our results could be affected by resolution conservatively by $50\%$ at the most (see Appendix of Paper I); we exclude this
region from the fitting procedures and from subsequent analyses.

\begin{table}
\begin{center}
\begin{tabular}{cccccc}
\hline
Redshift  & $q_X({\rm Mpc}/h)$ & $\alpha_X$ & $q_V({\rm km/s})$ & $\alpha_V$ & $\beta$\\
\hline
\hline
0.0    &   $11.82$  & $-0.4$ & $4.5\times10^4$    & $-0.33$ & $0.75$                                 \\  \hline
\hline
\end{tabular}
\end{center}
\caption{Values of the fitting parameters in Eq.~(\ref{lame_2}) at $z=0$ over the virialized region of the Aq-A-2 halo. 
Together with Eqs.~(\ref{lame}) and (\ref{lame_2}), these parameters provide a good description of \psad~at small $(\Delta x, \Delta v)$ (see Fig.~\ref{Fig_comp_fit}).}
\label{table_fit_0} 
\end{table}

\section{Dark matter annihilation rate: substructure boost}\label{sec_rate}

The number of dark matter self-annihilation events per unit time is given
in terms of the phase space distribution at the phase space coordinates 
${\bf x},{\bf v}$ and ${\bf x}+{\bf \Delta x},{\bf v}+{\bf \Delta v}$: 
\begin{eqnarray}\label{rate}
  R_{\rm ann} &=& \frac{1}{2m_{\chi}^2}\lim_{\Delta x \to 0}\left[\int d^3{\bf \Delta v}\bigg(\int_{{\cal V}_6} d^3{\bf v} d^3{\bf x} \right.\nonumber\\
    & & (\sigma_{\rm ann} v) f({\bf x},{\bf v})f({\bf x}+{\bf\Delta x},{\bf v}+{\bf \Delta v})\bigg)\bigg]\nonumber\\
  &=& \frac{M_{{\cal V}_6}}{2m_{\chi}^2} \int d^3{\bf \Delta v} (\sigma_{\rm ann} v) \lim_{\Delta x \to 0} \Xi(\Delta x, \Delta v)
\end{eqnarray}
where $M_{{\cal V}_6}$ is the total dark matter mass within the phase space volume ${\cal V}_6$, 
$(\sigma_{\rm ann} v)$ is the product of the annihilation cross section and the relative velocity between pairs\footnote{Note that in our notation,
the relative velocity is $\Delta v$, thus, $(\sigma_{\rm ann} v)\equiv(\sigma_{\rm ann} \Delta v)$. Throughout our paper we choose to use the former since this
is the common choice in the literature.}, and we have used Eq.(\ref{coarse}) to introduce \psad. 

The annihilation rate in a region of spatial volume $V$ is typically written as:
\begin{equation}\label{rate_0}
  R_{\rm ann} = \frac{1}{2m_{\chi}^2}\int_V d^3{\bf x}\rho^2({\bf x})\langle\sigma_{\rm ann} v\rangle
\end{equation}
where $m_\chi$ is the dark matter particle mass, $\rho_\chi({\bf x})$ is the local dark matter density (that includes contribution from the
smooth dark matter distribution and from substructure), and $\langle\sigma_{\rm ann} v\rangle$ is an average of $(\sigma_{\rm ann} v)$ over the 
velocity distribution of the dark matter particles (typically assumed to be Maxwellian). Instead of separating the dark matter spatial and
velocity distributions, Eq.~(\ref{rate}) gives directly, without any assumptions, the annihilation rate as an integral over the 
relative velocity $\Delta v$ at the limit of zero spatial separation of \psad, henceforth abbreviated $P^2SAD^{\rm zero}$($\Xi^{\rm zero}$). 
Since in general $(\sigma_{\rm ann} v)$ is an arbitrary 
function of $\Delta v$, we can use Eq.~(\ref{rate}) to easily accommodate any velocity dependent annihilation
cross section. Notice also that Eq.~(\ref{rate}) adapts to the region of interest for the annihilation rate by simply using 
\psad~averaged over that region. 

The near universality of \psad~at small scales (i.e. small separations in phase space) makes the functional shape in Eqs.~(\ref{lame}-\ref{lame_2}) 
valid across regions of substantially different ambient density with only slight changes to the fitting parameters (see sections 3.3 and 3.4 of 
Paper I) and can therefore be straightforwardly applied in Eq.~(\ref{rate}) to calculate the annihilation rate in gravitationally bound
substructure either globally in the whole halo, or locally in a certain region of the halo. Since we also found that at small scales 
\psad~is nearly insensitive to the assembly history of a particular halo (see Fig. 10 of Paper I), the results we find for the 
particular initial conditions of Aq-A-2 are also a very good approximation for different initial conditions. 

The integral in Eq.~(\ref{rate}) can also be easily transformed into an integral over $P^2SAD^{\rm zero}$ using Eq.~(\ref{lame}) 
and a change of variables:
\begin{equation}\label{rate_psad}
  R_{\rm ann} = \frac{2\pi M_{{\cal V}_6}}{m_{\chi}^2}q_V^3\alpha_V \int_{\Xi^{\rm zero}_{\rm max}}^{\Xi^{\rm zero}_{\rm min}} d\Xi^{\rm zero} (\sigma_{\rm ann} v) [\Xi^{\rm zero}]^{3\alpha_V}
\end{equation} 

Notice that because of the limit $\Delta x\rightarrow 0$, the annihilation rate is at the end only sensitive to $\alpha_V$ and $q_V$. 
The limits of the integral over $P^2SAD^{\rm zero}$ ($\Xi^{\rm zero}_{\rm max(min)}$) correspond to the maximum and minimum 
values of the separation in velocity where substructures contribute. Note that we can only 
use Eqs.~(\ref{lame}-\ref{lame_2}) down to: 
\begin{equation}\label{xi_min}
  \Xi^{\rm zero}_{\rm min}=\frac{10^8{\rm M}_\odot}{{\rm Mpc}^3({\rm km/s})^3}~h^2,
\end{equation}
below which $P^2SAD^{\rm zero}$ falls off more rapidly than the power law in Eq.~(\ref{lame_2}), and more importantly, 
the smooth distribution of dark matter dominates \psad~in general (Paper I). We will then take this value as the 
transition between the smooth and subhaloes dominance of the annihilation rate.

To compute the contribution from the smooth component we substitute $\rho$ in Eq.~(\ref{rate_0}) by $\rho_E$, the spherically 
averaged Einasto density profile:
\begin{equation}\label{einasto}
  \rho(r)=\rho_{-2}{\rm exp}\left(\frac{-2}{\alpha_e}\left[\left(\frac{r}{r_{-2}}\right)^{\alpha_e}-1\right]\right)
\end{equation}
where $\rho_{-2}$ and $r_{-2}$ are the density and radius at the point where the logarithmic density slope is -2, and $\alpha_e$ is the Einasto
shape parameter. We take the parameter values from the fit to the Aq-A-2 halo given in \citet{Navarro_10}: 
$\alpha_e=0.163$, $\rho_{-2}=3.9\times10^{15}~{\rm M}_\odot/{\rm Mpc}^3$, $r_{-2}=15.27~{\rm kpc}$. We also assume that the
velocity distribution of dark matter particles in the smooth component is Maxwellian. The substructure boost is then
simply defined as:
\begin{equation}\label{boost}
  B_{{\cal V}_6}=\frac{R_{\rm ann}^{\rm subs}}{R_{\rm ann}^{\rm smooth}}=
  \frac{M_{{\cal V}_6} \int d^3{\bf \Delta v} (\sigma_{\rm ann} v) \lim_{\Delta x \to 0} \Xi^{\rm subs}(\Delta x, \Delta v)}
  {\int_V d^3{\bf x}\rho_E^2({\bf x})\langle\sigma_{\rm ann} v\rangle_{\rm MB}}
\end{equation}

As an example, let us take the case where $(\sigma_{\rm ann} v)={\rm const}$ and compute the total annihilation rate in resolved substructures at $z=0$
within a MW-size halo. Since $\alpha_V\sim-1/3$ (see Table \ref{table_fit_0}), according to Eq.~(\ref{rate_psad}) we have:
\begin{equation}\label{rate_cte}
  R_{\rm ann}^{\rm subs}\propto \frac{q_v^3}{3}~{\rm ln}\left(\frac{\Xi^{\rm zero}_{\rm max}}{\Xi^{\rm zero}_{\rm min}}\right)
\end{equation}
If we take the value of $\Xi^{\rm zero}_{\rm min}$ given in Eq.~(\ref{xi_min}) and the maximum value of \psad~we can resolve in Aq-A-2, 
$\Xi^{\rm zero}_{\rm max}\sim10^{12}{\rm M}_\odot h^2/({\rm Mpc~\rm km/s})^3$, then we can estimate the ``resolved'' substructure boost to the total 
annihilation rate in the Aq-A-2 halo:
\begin{equation}\label{boost_res}
  B_{\rm Aq-A-2}^{\rm res}\sim0.61
\end{equation}
The simulation particle mass of Aq-A-2 is $m_p=1.37\times10^4$M$_\odot$ and the minimum subhalo mass that we can trust in terms of global 
subhalo properties is $\sim10^6$M$_{\rm \odot}$ ($\sim75$ particles; see Figs. 26 and 27 of \citealt{Springel_08}). By using a subhalo model
(see Section 3.2 of Paper I), we have checked that this minimum ``resolved'' subhalo mass also corresponds roughly to the value of 
$\Xi^{\rm zero}_{\rm max}$ we can resolve. As a consistency check, we can therefore compare the value in Eq.~(\ref{boost_res}) to the subhalo boost 
computed in \citet{Springel_08b} (using the same simulation data)
by summing up the contribution of all resolved subhaloes above $10^6$M$_{\rm \odot}$ within the virial radius of the Aq-A-1 halo, and 
assuming a NFW density profile for each of them (see second green line from top to bottom in Fig. 3 of \citealt{Springel_08b}). The resolved
subhalo boost they found is $\sim0.5$, quite similar to ours.

Before continuing with the analysis of the applications of \psad~to compute the dark matter annihilation rate, we describe in the next section
a physically-motivated model that we will use to extrapolate the behaviour of \psad~to the unresolved regime.

\section{Improved stable clustering hypothesis and spherical collapse model}\label{stable_model}

The stable clustering hypothesis was originally introduced by \citet{Davis_77} to study the galaxy two-point correlation
function in the strongly non-linear regime. The hypothesis proposes that the number of neighbouring dark matter particles 
within a fixed physical
separation becomes constant (i.e. there is no net streaming motion between particles in physical coordinates) on sufficiently
small scales. \citet{Jing_01} found that the hypothesis is valid when averaging the mean pair velocity between particles in
many simulated haloes but found that this is not generally satisfied within one single virialized halo. 

The hypothesis can
be extended to phase space \citep{Afshordi_10} by stating that, on average, the number of particles within the physical distance
${\bf \Delta x}$ and physical velocity ${\bf \Delta v}$ of a given particle does not change with time for sufficiently
small ${\bf \Delta x}$ and ${\bf \Delta v}$. In Paper I we found evidence for the validity of the
stable clustering hypothesis in phase space through the analysis of \psad~at small scales~finding that it typically varies
by a factor of a few in regions of substantially different ambient densities (by nearly four orders of magnitude).

The small scale structure of \psad~today within the virialized region of the halo is given by a collection
of gravitationally bound merging substructures that collapsed earlier than the host halo\footnote{From here onwards, we will 
follow closely the notation given in \citet{Afshordi_10}.}. Each of these
substructures has a characteristic phase space density $\xi_s$ imprinted at the collapse time, when the Hubble's constant has a value
$H(\xi_s)$, and each has a collapse mass $m_{\rm col}(\xi_s)$. In the absence of tidal disruption, $\xi_s$ would be preserved from the
time of collapse until today. The stable clustering hypothesis can then be used and applied to write a simplified solution
to the collisionless Boltzmann equation at the phase space coordinates $({\bf \Delta x},{\bf \Delta v})$
of the following form (for a spherically symmetric gravitational potential, \citealt{Afshordi_10}):
\begin{eqnarray}\label{ellipse_0}
  \Xi^{\rm sub}(\Delta x, \Delta v) &\equiv& \mu(m_{\rm col})\xi_s\nonumber\\ 
  &=& F\left[({\bf \Delta v})^2 + (4\pi/3)G\rho_{\rm char}({\bf \Delta x})^2\right]
\end{eqnarray}
where $\rho_{\rm char}$ is the characteristic density of the collapsed subhalo, which is roughly $\sim200$ times the critical density
at the collapse time. Since we know that subhaloes will be subjected to tidal disruption as they merge and move through denser
environments, Eq.~(\ref{ellipse_0}) accounts for tidal stripping modifying the stable clustering hypothesis
prediction (i.e. $\xi_s=F$) by introducing $\mu(m_{\rm col})$, which can be interpreted as the mean fraction of particles that remain bound. 
To make the connection with the simulation results, we equate Eq.~(\ref{coarse}) (in the small-scale regime) with Eq.~(\ref{ellipse_0}). 
We can then associate a constant value of $\Xi$ within the simulated MW-size halo to a subhalo that formed with a typical mass $m_{\rm col}$ in the past. 

The characteristic phase space density $\xi_s$ of a given (sub)halo at formation time can be estimated
within the spherical collapse model using the characteristic densities and velocities
of the collapsed object: $\rho_{\rm char}\equiv200\rho_{\rm crit}$ and $\sigma_{\rm char}\equiv\sigma_{\rm vir}=10Hr_{\rm vir}$ 
\citep[e.g.][]{Afshordi_02}:
\begin{equation}\label{rho_ps}
\xi_s=\frac{\rho_{\rm char}}{\sigma_{\rm vir}^3}=\frac{10H(\xi_s)}{G^2m_{\rm col}(\xi_s)}.
\end{equation}
In the spherical collapse model, the subhalo collapses when the r.m.s top-hat linear overdensity $\sigma(m_{\rm col})$ (mass variance)
crosses the linear overdensity threshold $\delta_c\sim1.7$ at an epoch given by:
\begin{equation}\label{hubble}
H(\xi_s)\sim H_0\left(\frac{\sigma(m_{\rm col})}{\delta_c}\right)^{3/2}
\end{equation}
The mass variance is defined by:
\begin{equation}
\sigma^2(m_{\rm col})\equiv\frac{1}{(2\pi)^3}\int_0^\infty 4\pi k^2P(k)W^2(k;m_{\rm col})dk
\end{equation}
where $W(k;m_{\rm col})$ is the top-hat window function and $P(k)$ is the linear power spectrum. We use the fitting function given in 
\citet{Taruya_00} which is accurate to a few percent in the mass range we use here:
\begin{equation}
\sigma(m_{\rm col})\propto\left( 1 + 2.208m^p - 0.7668m^{2p} + 0.7949m^{3p} \right)^{-2/(9p)},
\end{equation}
where $p=0.0873$ and $m=m_{\rm col}(\Gamma h)^2/10^{12}$M$_{\odot}$, where $\Gamma=\Omega_mh~{\rm exp}(\Omega_b-\sqrt{2h}\Omega_b/\Omega_m)$
with $\Omega_m$ and $\Omega_b$ being the contributions from matter and baryons to the mass energy density of the Universe. The 
mass variance is normalized to the value at $8h^{-1}$~Mpc spheres at redshift zero. We assume the same cosmological parameters as
in the Aquarius simulations (those of a WMAP1 flat cosmology): $\Omega_m=0.25$, $\Omega_\Lambda=1-\Omega_m$, $h=0.73$, $\sigma_8=0.9$ 
and $n_s=1$ (the spectral index of the primordial power spectrum). 

Using Eq.~(\ref{hubble}) we can then calculate the epoch of collapse of
a given halo of mass $m_{\rm col}$ and estimate its phase space density $\xi_s$ using Eq.~(\ref{rho_ps}). Since the area enclosed by the
ellipse: $({\bf \Delta v})^2 + 100H^2({\bf \Delta x})^2 = F^{-1}$ is simply $\pi F^{-1}/10H$ then 
we can equate the phase space volume encompassed by the ellipsoid to the volume of the collapsed halo $m_{\rm col}/\xi_s$:
\begin{equation}\label{F_1}
\left(\frac{\pi F^{-1}(\mu\xi_s)}{10H(m_{\rm col})}\right)^3=\frac{G^2m_{\rm col}^2}{10H(m_{\rm col})}
\end{equation}
Thus, we can finally give the prediction of the improved stable clustering hypothesis for the curves of contours of constant \psad:
\begin{equation}\label{stable_pred}
\left(\frac{\Delta x}{\lambda(m_{\rm col})}\right)^2 + \left(\frac{\Delta v}{\zeta(m_{\rm col})}\right)^2 = 1 
\end{equation}
where $\lambda^2(m_{\rm col})=F^{-1}/100H^2$ and $\zeta^2(m_{\rm col})=F^{-1}$.

Comparing this prediction with the simulation data, we find that ellipses are not a good description, 
instead generalizing the ellipses to superellipses (Lam\'e curves) provides a good fit to the simulated MW halo 
at small $(\Delta x, \Delta v)$ (i.e., to $\Xi^{\rm sub}(\Delta x, \Delta v)$):
\begin{equation}\label{lame_mod}
\left(\frac{\Delta x}{a\lambda(m_{\rm col})}\right)^{\beta} + \left(\frac{\Delta v}{b\zeta(m_{\rm col})}\right)^{\beta} = 1 
\end{equation}
Also, we find that a better fit to the simulated data is obtained from a mass dependent tidal disruption parameter 
$\mu$ instead of the constant value taken by \citet{Afshordi_10}. In what follows, we introduce a model for tidal 
disruption that captures this mass dependence.

\subsection{Tidal stripping}\label{tidal_model}

We introduce a simplified model of tidal stripping in which we assume that the characteristic 
density in a given subhalo changes with time according to:
\begin{equation}\label{model_st_1}
  \frac{{\rm d}\rho_{\rm sub}}{{\rm d}t}=-\frac{\rho_{\rm sub}}{\tau_{\rm ff}}F_{\rm tid}\left(\frac{\rho_{\rm sub}}{\rho_{\rm host}}\right),
\end{equation}
where $\tau_{\rm ff}\propto1/\sqrt{G\rho_{\rm host}}$ is the characteristic free-fall time of the host as the subhalo starts being stripped. 
If we assume that $F_{\rm tid}=A_{\rm tid}\left(\rho_{\rm sub}/\rho_{\rm host}\right)^{-\alpha}$, then the solution 
to Eq.~(\ref{model_st_1}) as a function of redshift is:
\begin{eqnarray}\label{model_st_2}
  \left(\frac{\rho_{\rm sub}(z)}{\rho_{\rm sub}(z_{\rm inf})}\right)^{\alpha}&=& 
  1 - \frac{\alpha A_{\rm tid}\sqrt{75/\pi}}{\left(\Omega_m\left(1+z_{\rm inf}\right)^3+\Omega_\Lambda\right)^\alpha}\nonumber\\
  &\times&G(z,z_{inf};\alpha),
\end{eqnarray}
where
\begin{equation}\label{g_aux}
  G(z,z_{inf};\alpha) = \int_z^{z_{\rm inf}}\frac{\left(\Omega_m\left(1+z\right)^3+\Omega_\Lambda\right)^\alpha}{1+z}dz,
\end{equation}
with $z_{\rm inf}$ being the relevant redshift of first infall and we have assumed that 
$\rho_{\rm host}(z)=200\rho_{\rm crit}(z)$ and $\rho_{\rm sub}(z_{\rm inf})=200\rho_{\rm crit}(z_{\rm inf})$.
If we take as an ansatz that $\mu$ evolves in a similar way as $\rho_{\rm sub}$, then we can write:
\begin{equation}\label{mu_final}
  \mu(z) = \mu_0(z_{\rm inf};m_{\rm col})\left(\frac{\rho_{\rm sub}(z)}{\rho_{\rm sub}(z_{\rm inf})}\right).
\end{equation}
The value of $\mu$ at the time of first infall is uncertain, but considering that structures that formed earlier would be more resilient to 
tidal stripping, and that tidal disruption begins as the subhalo infalls into a larger structure (not necessary the final 
host halo) with a mass $fm_{\rm col}$ ($f>1$), we model the initial condition as:
\begin{equation}
  \mu_0(z_{\rm inf};m_{\rm col})^\alpha=\tilde{B}\left[\frac{\sigma(m_{\rm col})}{\sigma(fm_{\rm col})}\right]^{2\kappa},
\end{equation}
where $\tilde{B}$ and $\kappa$ are free parameters.

The infall redshift can be estimated using the Extended Press-Schechter formalism. We are interested in the probability that a halo 
of mass $m_{2}(z_{\rm inf})=fm_{\rm col}(z_{\rm col})$, had progenitors of mass $m_{\rm col}(z_{\rm col})$: 
\begin{eqnarray}
  P[\delta_{c}(1+z_{\rm col})]\propto {\rm exp}
  \left[\frac{-\left(\delta_c(1+z_{\rm col})-\delta_c(1+z_{\rm inf})\right)^2}{2\left(\sigma^2(m_{\rm col})-\sigma^2(fm_{\rm col})\right)}\right]\nonumber\\
  \times{\rm exp}\left[\frac{-\delta_c^2(1+z_{\rm inf})}{2\sigma^2(fm_{\rm col})}\right],
\end{eqnarray}
where $\delta_c(1+z)\propto(1+z)$ (at early times) is the overdensity barrier required for spherical collapse. Using the method of steepest 
descent, we can approximate the average infall time:
\begin{equation}
1+z_{\rm inf}\approx\left[\frac{\sigma(fm_{\rm col})}{\sigma(m_{\rm col})}\right]^2(1+z_{\rm col}).
\end{equation}
Since $\sigma(m_{\rm col})\equiv\delta_c(z_{\rm col})\approx1.7/D(z_{\rm col})$, where $D(z)$ is the linear growth factor 
\citep[for an approximation formula see e.g.][]{Carroll_92}, we can then obtain
the collapse redshift from the mass variance and finally obtain the value of $\mu(z)$ for a given $m_{\rm col}$.

We introduce a halo-centric distance dependence in $\mu$ by considering that the average density of the host within a given distance $r$ 
was established at an epoch $z^{\ast}$ when its characteristic density had that value: $\rho_{\rm host}(<r)=200\rho_{\rm crit}(z^{\ast})$ (this
defines the characteristic free fall time). We also use the radial diagonal part of the Hessian of the potential $\phi(r)_{,rr}$ as the quantity that
drives tidal disruption, rather than simply $\rho_{\rm host}(r)$. 

\begin{figure*}
\begin{tabular}{|@{}l@{}|@{}l@{}|}
\includegraphics[height=8.0cm,width=8.0cm]{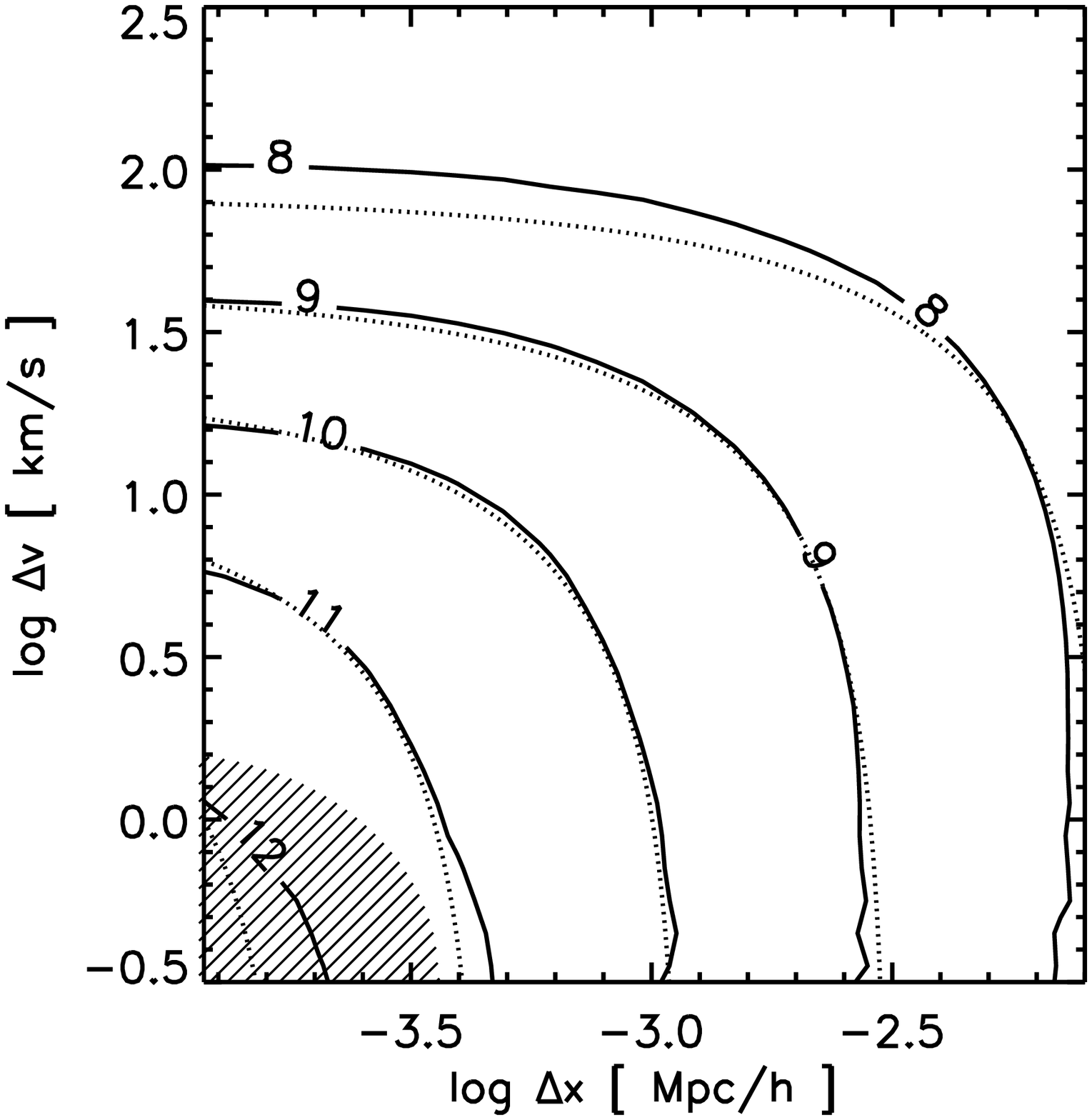} &
\includegraphics[height=8.0cm,width=8.0cm]{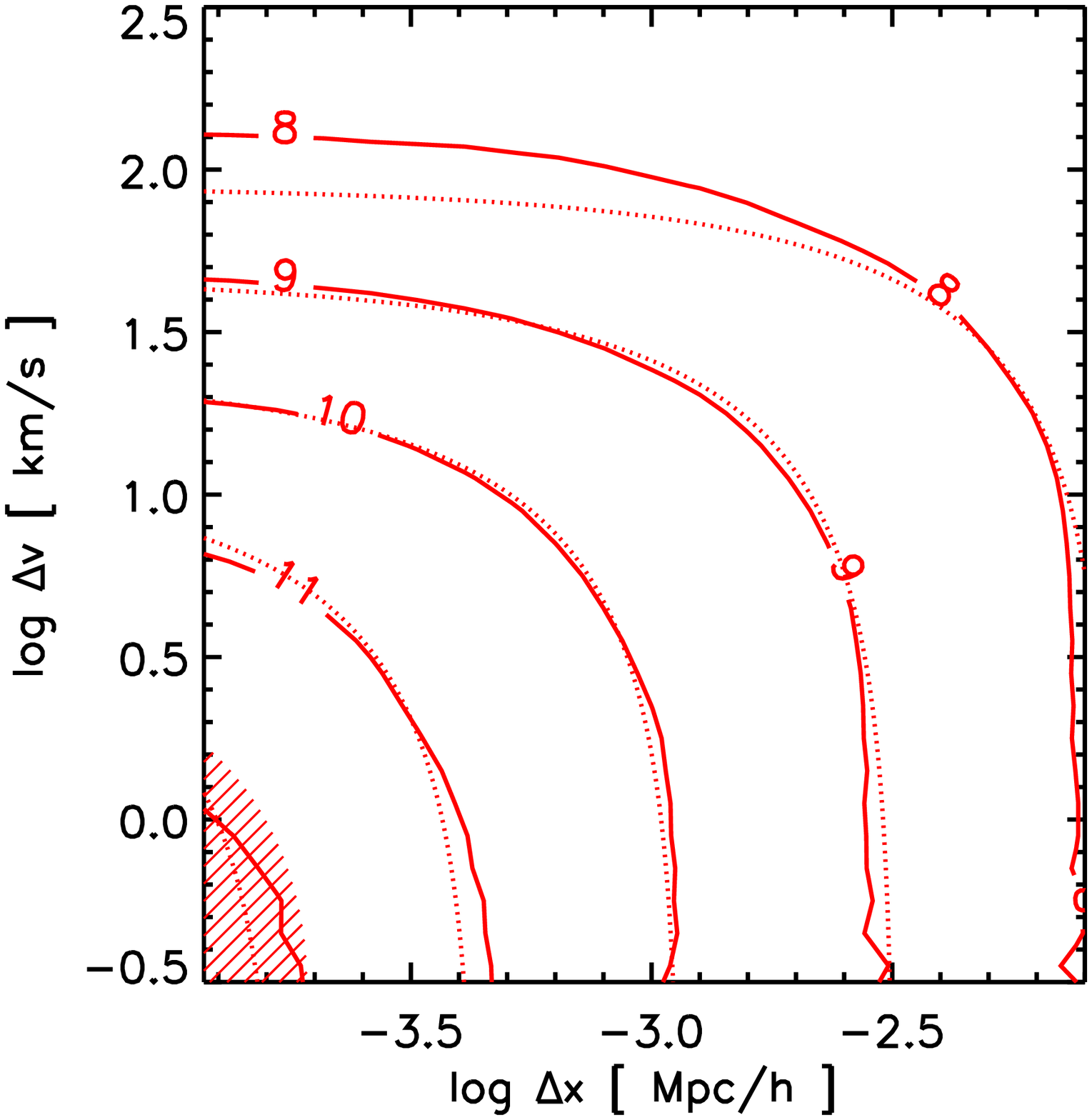} \\
\includegraphics[height=8.0cm,width=8.0cm]{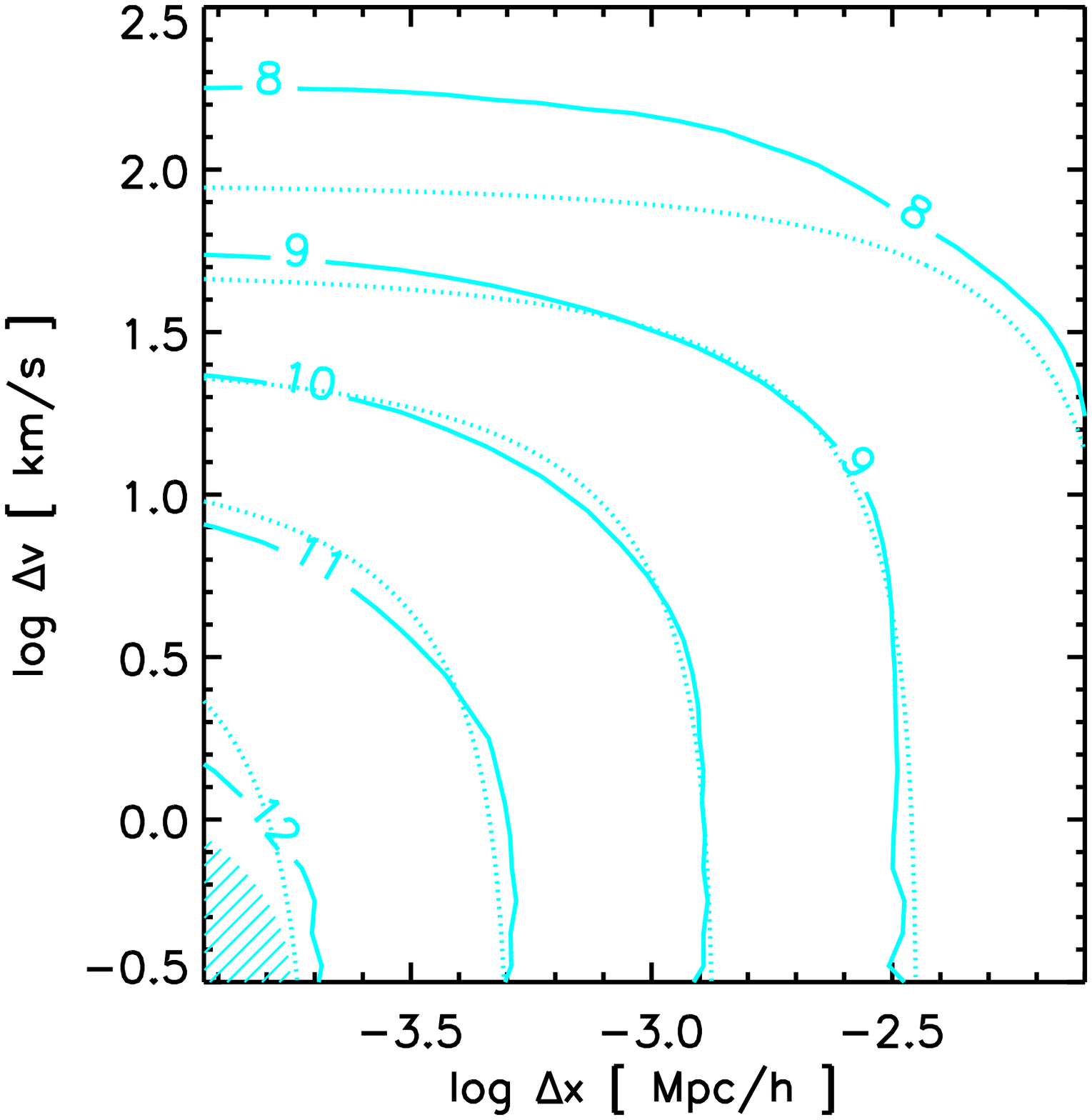} &
\includegraphics[height=8.0cm,width=8.0cm]{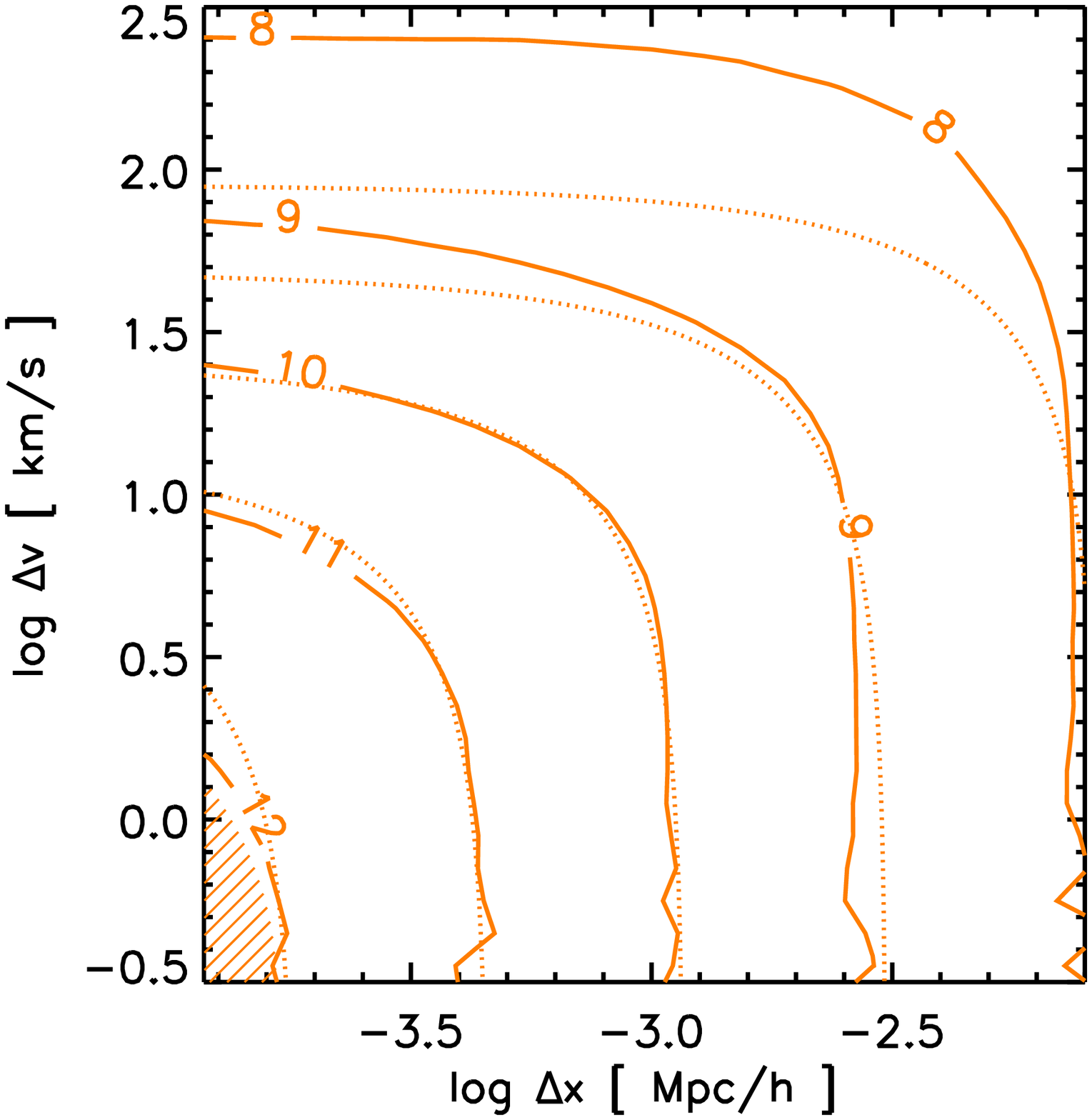} \\
\end{tabular}
\caption{Contours of the logarithm of the particle phase space average density (\psad) of the Aq-A-2 halo (solid) and for the
model described in Section \ref{stable_model} (dotted)  at different redshifts, clockwise from top left: 
$z=0$ (black), $z=0.95$ (red), $z=2.2$ (cyan) and $z=3.5$ (orange). The dashed regions in the left corners encompass the regions where 
resolution is a potential issue (see Appendix of Paper I).}
\label{Fig_comp_stable_red} 
\end{figure*}

After $z^\ast$, the density does not dilute anymore as the Universe expands. 
If $z_{\rm inf}>z^{\ast}$ we then have an additional solution to Eq.~(\ref{model_st_1}) for $z<z^{\ast}$:
\begin{eqnarray}\label{model_st_3}
  \left(\frac{\rho_{\rm sub}(z;r)}{\rho_{\rm sub}(z^{\ast})}\right)^{\alpha}&=& 
  1 - \alpha A_{\rm tid}\sqrt{75/\pi}\left.\left[\frac{\rho_{\rm host}(<r)}{200\rho_{\rm crit}(z=0)}\right]^{1/2}\right\vert_{\rm ren}\nonumber\\
  &\times&\left.\left(\frac{\phi(r)_{,rr}}{4\pi G\rho_{\rm host}(r)}\right)^\alpha\right\vert_{\rm ren} T(z^{\ast},z),
\end{eqnarray}
where
\begin{equation}
  T(z^\ast,z) = \int_z^{z^\ast}\frac{\left(\Omega_m\left(1+z\right)^3+\Omega_\Lambda\right)^{-1/2}}{1+z}dz,
\end{equation}
and the subscript $ren$ means that the quantity is renormalized to the radius where the average solution applies, i.e., to $r=r_{200}$. If
$z_{\rm inf}<z^{\ast}$, then only this ``second mode'' (Eq.\ref{model_st_3}) of stripping occurs.

In the end, we have a model of tidal disruption with $5$ free parameters: $A_{\rm tid}$, $\alpha$, $f$, $\tilde{B}$ and $\kappa$, in addition 
to the improved stable clustering prediction which includes $3$ additional free parameters (which in principle could depend on 
redshift and in 
halo-centric distance): $a(z,r)$, $b(z,r)$ and $\beta(z,r)$ accounting for the stretching in phase space due to tidal shocking 
(Eq.\ref{lame_mod}). This model is able to describe the small-scale behaviour of \psad~at different redshifts and at different 
regions within the halo as we show in the following section.

\section{Simulation data and model comparison}\label{sim_fit}

\subsection{Average behaviour within the virialized halo and redshift dependence}

We first compare the model developed in Section \ref{stable_model} with the value of \psad~averaged over all particles that are 
gravitationally bound to the Aq-A-2 halo, i.e., those in the smooth and substructure components (for a more detailed definition see the first paragraphs
of section 3 of Paper I). Fig.~\ref{Fig_comp_stable_red} shows the small scale behaviour of \psad~in the simulation at  different
redshifts (solid lines) and fits by our model in dotted lines. Although the fits are poorer for $\Xi>10^9{\rm M}_\odot h^2/({\rm Mpc~\rm km/s})^3$ 
(particularly at $z>2$), it is clearly a reasonable description at smaller scales (i.e., separations in phase space), which are the ones that 
matter the most to extrapolate the behaviour to the unresolved scales. 

The variations across redshift can be accommodated by introducing a slight redshift dependence of the parameters $\beta$, $a$ and $b$. The 
former is given by a simple linear relation as in Paper I:
\begin{equation}\label{beta_z}
  \beta(z)=0.67+0.08(1+z)
\end{equation}
while $a(z)$ and $b(z)$ are given in Table \ref{table_fit}.

\begin{table}
\begin{center}
\begin{tabular}{ccc}
\hline
Redshift  & $a$ & $b$ \\
\hline
\hline
0.0    &   $0.29$  & $1.77$                                    \\  \hline
0.95   &   $0.26$  & $1.75$                                    \\  \hline
2.2    &   $0.28$  & $1.77$                                   \\  \hline
3.5    &   $0.24$  & $1.77$                                    \\  \hline
\hline
\end{tabular}
\end{center}
\caption{Values of the fitting parameters in Eq.~(\ref{lame_mod}) that together with the tidal disruption model explained in Section
\ref{tidal_model} provide a good description of \psad~at small $(\Delta x, \Delta v)$ (see Fig.~\ref{Fig_comp_stable_red}).}
\label{table_fit} 
\end{table}

The full tidal stripping model described in Section \ref{tidal_model} fits the simulated data with the following values for the five free parameters: 
$f=2.1$, $\alpha=1/3$, $A_{\rm tid}=0.12$, $\tilde{B}=0.185$ and $\kappa=4.5$. Although we did not explore exhaustively the large parameter
space we found that large deviations from these values seem to give a poorer fit (this is especially the case for $f$ and $\alpha$). We also note
that the behaviour of $\mu$ as a function of mass and redshift can be roughly approximated by the following
formula (up to $z\sim2$): 
\begin{equation}
  \mu(z;m_{\rm col})\approx0.01\left[1+\left(\frac{m_{\rm col}(1+z)^{2.5}}{10^8{\rm M}_{\odot}}\right)^{0.22}\right]
\end{equation}

\subsection{Halo-centric distance dependence}\label{hc_dist_sec}

\begin{figure*}
\begin{tabular}{|@{}l@{}|@{}l@{}|}
\includegraphics[height=8.0cm,width=8.0cm]{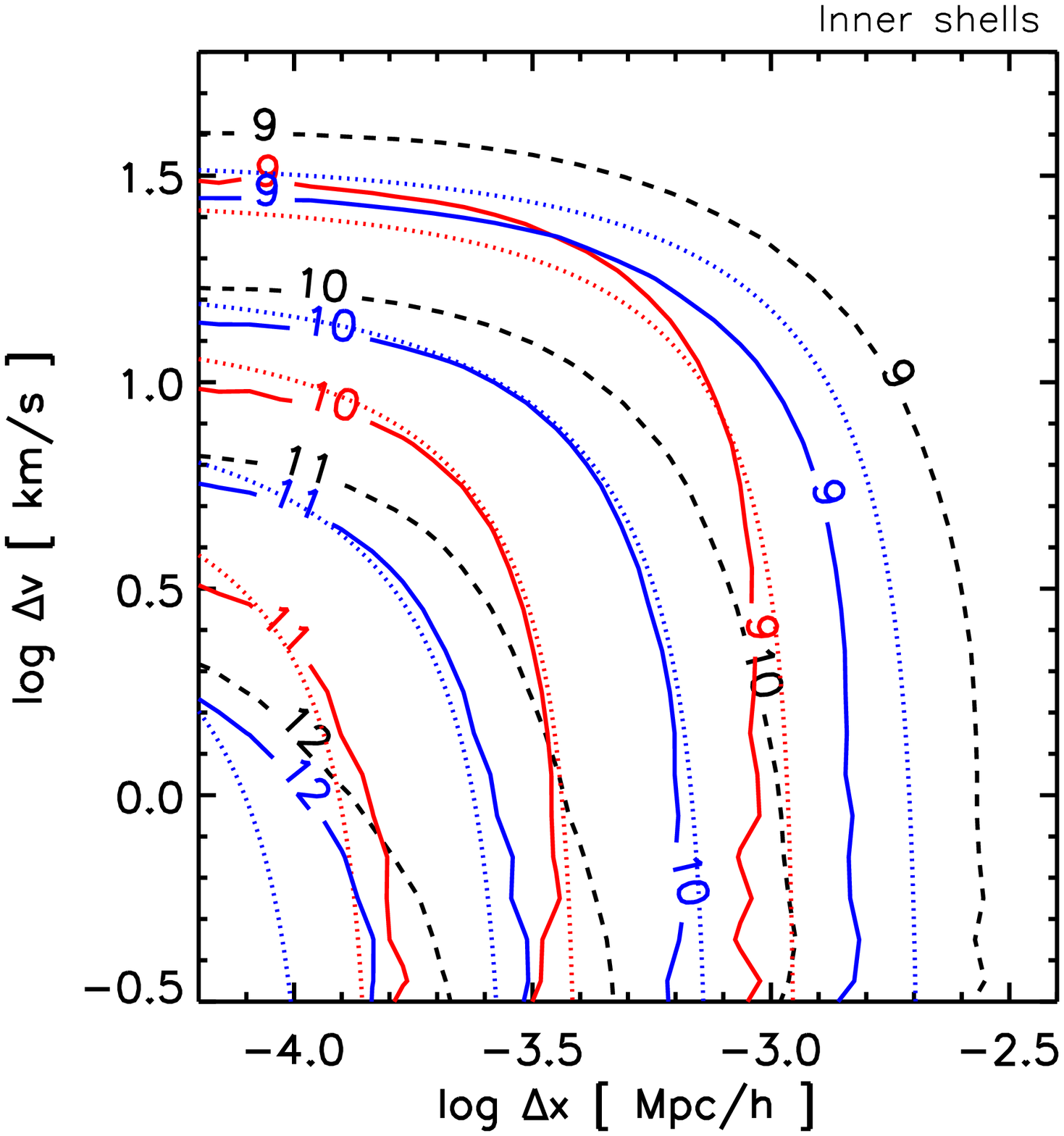} &
\includegraphics[height=8.0cm,width=8.0cm]{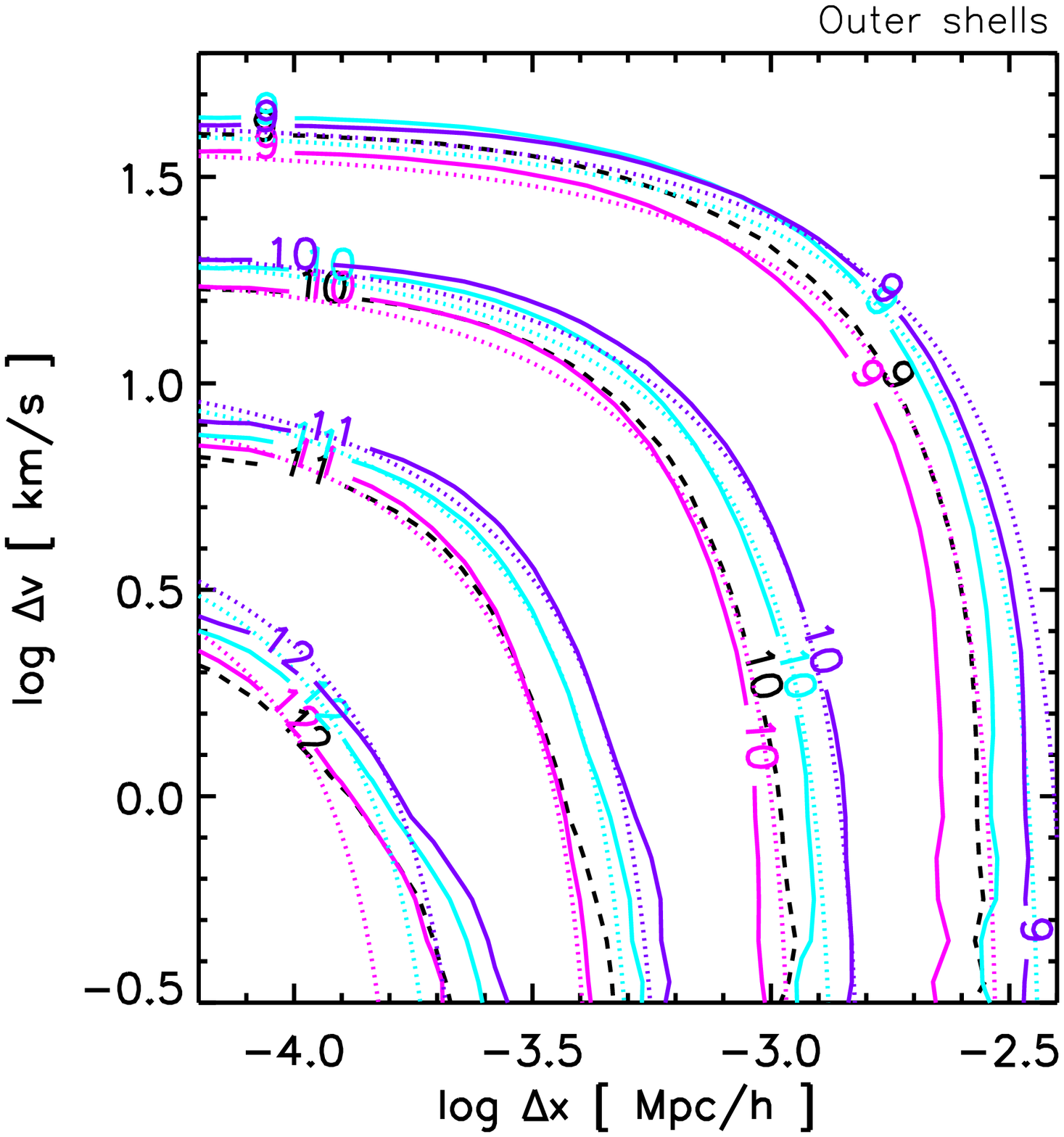} \\
\end{tabular}
\caption{Contours of the logarithm of the particle phase space average density (\psad) for the Aq-A-2 halo 
at $z=0$ for samples of particles taken at different radial shells from the halo centre. {\it Left panel}:
$0<r/r_{200}<0.2$ (red) and $0.2<r/r_{200}<0.4$ (blue). {\it Right panel}: $0.4<r/r_{200}<0.6$ (magenta), $0.6<r/r_{200}<0.8$ 
(cyan), and $0.8<r/r_{200}<1.0$ (violet). The dotted lines are fits given by our model described in section \ref{stable_model}. The
relevant parameters of the fits are given in Table \ref{table_fit_radial}. For reference, the dashed line is \psad~averaged over the whole halo.} 
\label{Fig_radial_dep} 
\end{figure*}

Fig.~\ref{Fig_radial_dep} shows the contours of log(\psad) averaged over particles that are located in concentric shells at different distances 
from the halo centre as described in the caption. The solid lines are the simulation data for the Aq-A-2 halo while the dashed lines are the result 
of our model with the same value of the free parameters associated to $\mu$ as in the previous average case (i.e., $f=2.1$, $\alpha=1/3$, 
$A_{\rm tid}=0.12$, $\tilde{B}=0.185$ and $\kappa=4.5$). We note that $\mu$ is a function of radius through Eq.~(\ref{model_st_3}) that can be
approximated by:
\begin{equation}\label{mu_r}
  \mu(r;m_{\rm col})\approx\mu_{0}(r)\left[c_\mu(r)+\left(\frac{m_{\rm col}}{10^8{\rm M}_{\odot}}\right)^{\alpha_\mu(r)}\right]
\end{equation}
where the radial-dependent parameters $\mu_0(r)$, $c_{\mu}(r)$ and $\alpha_\mu(r)$, as well as the shape parameters $a$, $b$ and $\beta$, that fit the 
simulation data are given in Table \ref{table_fit_radial}. 
\begin{table}
\begin{center}
\begin{tabular}{ccccccc}
\hline
$r/r_{200}$   & $\beta$ & $a$ & $b$ & $\mu_0$ & $\alpha_\mu$ & $c_{\mu}$\\
\hline
\hline
0.0-0.2    & $1.0$   &   $0.17$   & $1.55$ & $0.0045$ & $0.15$ & $0.15$                                \\  \hline
0.2-0.4    & $0.9$   &   $0.20$   & $1.45$ & $0.0135$ & $0.13$ & $0.20$                                  \\  \hline
0.4-0.6    & $0.8$  &   $0.28$  & $1.55$ & $0.0160$ & $0.12$ & $0.25$                                   \\  \hline
0.6-0.8    & $0.775$ &   $0.35$   & $1.72$ & $0.0160$ & $0.12$ & $0.25$                                 \\  \hline
0.8-1.0    & $0.75$   &   $0.40$   & $1.80$ & $0.0160$ & $0.12$ & $0.25$                                 \\  \hline
\hline
\end{tabular}
\end{center}
\caption{The second, third and fourth columns are the values of the parameters in Eq.~(\ref{lame_2}) for the fit to \psad~averaged over different 
radial shells 
centered in the Aq-A-2 halo as indicated in the first column (see Fig.\ref{Fig_radial_dep}). The last three columns are the parameters of a simple
approximation (Eq.~\ref{mu_r}) of the parameter that controls tidal stripping ($\mu(r;m_{\rm col})$; see Section \ref{tidal_model}).}
\label{table_fit_radial} 
\end{table}
\section{Boost to Dark Matter Annihilation due to unresolved sub-structure}\label{boost_sec}

Once we have calibrated the model to the simulation data, we can use it to extrapolate the behaviour of \psad~to scales that are unresolved. 
We argue that this extrapolation method offer advantages over other commonly used methods in the computation of the boost to 
the annihilation rate due to gravitationally bound unresolved substructures. This is the case especially for methods that extrapolate the 
concentration-mass relation. The support for this argument is two-fold:
\begin{itemize}
  \item The small scale modelling of \psad~is physically motivated by: the stable clustering hypothesis in phase space, the spherical collapse model,
    and a simplified tidal disruption description.
  \item The structure of \psad~is directly connected to the annihilation rate (Eq.\ref{rate}), with the small scale behaviour reflecting the 
    substructure contribution. 
\end{itemize}
By calibrating the model to \psad, we avoid the use of the subhalo model and instead of uncertainties on the abundance of subhaloes, their radial 
distribution, and their internal properties, our model is sensitive to the behaviour of $\mu(m_{\rm col})$ and to the assumption that 
all other shape parameters
remain unchanged at unresolved scales. However, as we show below, of all these, $b$ is the only one of relevance.

As a first example, let us compute the total annihilation rate due to substructures within a MW-size halo at $z=0$ in the case of 
$(\sigma_{\rm ann} v)={\rm const}$:

\begin{widetext}
\begin{eqnarray}\label{rate_cte_model}
  R_{\rm ann}^{\rm subs}&=&\frac{4\pi M_{200}(\sigma_{\rm ann} v)}{2m_{\chi}^2} \int (\Delta v)^2d\Delta v \lim_{\Delta x \to 0} \Xi^{\rm sub}(\Delta x, \Delta v)
  =\frac{4\pi M_{200}(\sigma_{\rm ann} v)}{2m_{\chi}^2}\int \frac{b^3}{2}(F^{-1}(\mu\xi_s))^{1/2}\frac{dF^{-1}}{dm_{\rm col}}
  \mu(m_{\rm col})\xi_s(m_{\rm col})dm_{\rm col}\nonumber\\
  &=&\frac{8\pi^{1/2}b^3}{9\delta_c^3}200\rho_{crit,0}M_{200}\frac{(\sigma_{\rm ann} v)}{2m_\chi^2}
  \int_{\rm m_{min}}^{\rm m_{max}} \mu(m_{\rm col})m_{\rm col}^{-2} d(m_{\rm col}^2\sigma^3(m_{\rm col}))
\end{eqnarray}
\end{widetext}
\noindent where we have used Eqs.(\ref{ellipse_0}),(\ref{hubble}),(\ref{F_1}), and (\ref{lame_mod}). Notice that only $b$ and $\mu$ enter into
the annihilation rate (i.e., $\beta$ and $a$ are irrelevant). The limits of the integral are the minimum and maximum subhalo masses {\it at collapse} that
contribute to \psad, which are non-trivially connected to the tidally disrupted subhalo masses today. With the values of $\Xi^{\rm zero}_{\rm min}$ ($\Xi^{\rm zero}_{\rm max}$)
mentioned in Section \ref{sec_rate} corresponding to resolved subhaloes we can directly estimate $m_{\rm min}\sim10^7$M$_\odot$ 
($m_{\rm max}\sim10^{11}$M$_\odot$) since $\Xi=\mu\xi_s$. These masses are a factor of several above the minimum (maximum) subhalo masses at $z=0$
contributing to \psad~in the Aq-A-2 halo.

\begin{figure}
\center{
\includegraphics[height=8.0cm,width=8.0cm]{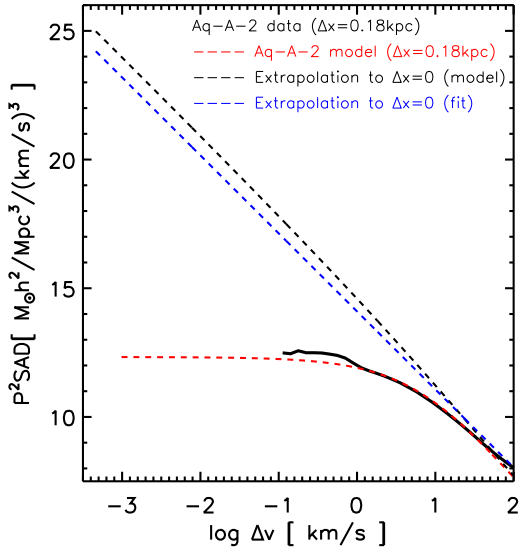}
}
\caption{Projection of the particle phase space average density (\psad) for the Aq-A-2 halo 
at $z=0$ in the $\Delta v$ direction fixing $\Delta x\sim0.18$~kpc (black solid). The red dashed line is the fit with our model using
the same cut in $\Delta x$, while the black and blue dashed lines are extrapolations to $\Delta x=0$ using our full model and using
the fitting function in Eq.~(\ref{lame}), respectively. The horizontal axis is extended all the way to the value of $\Delta v$ corresponding
to $m_{\rm col}=10^{-6}$M$_\odot$, the typical mass of WIMP models.} 
\label{Fig_projection} 
\end{figure}

The projection of \psad~along the $\Delta v$ direction can be seen in Fig.~\ref{Fig_projection} for the whole virialized region of Aq-A-2 at $z=0$. 
This is the quantity of interest to compute the annihilation rate and we show with a black solid line the case when $\Delta x\sim0.18$~kpc (which
is the minimum physical separation we can resolve). The dashed red line is a fit by our full model with the same cut in $\Delta x$ while the
limit to $\Delta x=0$ in the model is shown with a black dashed line. The extrapolation using the fitting function that directly relates $\Delta v$ with
\psad~(see Eq.~\ref{lame}) is shown with a blue dashed line. We have extended the horizontal axis to the corresponding velocity separations of
WIMP models with $m_{\rm col}=10^{-6}$M$_\odot$. From Fig.~\ref{Fig_projection} we can see that the simple formula (see Eq.~\ref{lame}):
\begin{equation}
  \lim_{\Delta x \to 0} \Xi^{\rm sub}(\Delta x, \Delta v) = (\Delta v/q_V)^{3}
\end{equation}
provides a good approximation to our full modelling and it immediately suggests that the annihilation rate scales roughly logarithmically with 
\psad~(for $\sigma_{\rm ann} v={\rm const}$).

\begin{figure}
\center{
\includegraphics[height=8.0cm,width=8.0cm]{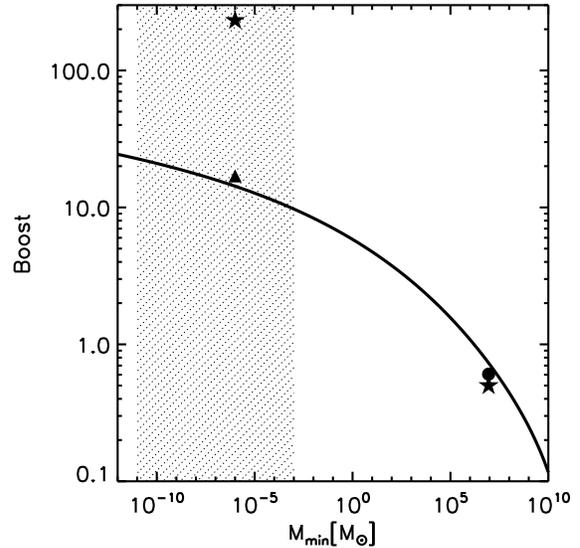}
}
\caption{Total subhalo boost to the annihilation rate (assuming $(\sigma_{\rm ann} v)={\rm const}$) within the virialized region of a MW-size halo as a 
function of the minimum subhalo mass {\it at collapse}. The filled circle shows the value of the boost estimated directly from the fit to the 
simulation data (Eq.~\ref{boost_res}), while
the star symbols are the values in the resolved ($m_{\rm sub}(z=0)\geq10^{6}$M$_{\odot}$) and extrapolated cases ($m_{\rm sub}(z=0)\geq10^{-6}$M$_{\odot}$) taken
from \citet{Springel_08b}. The filled triangle is the boost estimated by \citet{Kamionkowski_10} using a method calibrated with the Via Lactea II 
simulation \citep{Diemand_08}. The dotted region shows the theoretical expectation for the decoupling mass 
limit of WIMPs \citep{Bringmann_09}.} 
\label{Fig_boost} 
\end{figure}

We use Eq.~(\ref{rate_cte_model}) to compute the subhalo boost (using the smooth Einasto distribution described in Section \ref{sec_rate}) 
as a function of the minimum collapsed mass down to the decoupling masses corresponding to WIMPs; this is shown in Fig.~\ref{Fig_boost}. The filled circle
is the value in Eq.~(\ref{boost_res}) corresponding to resolved substructures and found directly with the small scale fitting function of \psad~in 
Section \ref{sec_rate} (i.e., Eq.~\ref{rate_cte}). The star symbol on the right (left) is shown for reference, and corresponds to the subhalo boost 
reported by 
\citet{Springel_08b} for a minimum subhalo mass $m_{\rm sub}(z=0)=10^{6}$($10^{-6}$)~M$_{\odot}$. The extrapolation in this case was done
under the assumption that the radial subhalo luminosity profile preserves its shape and that the re-scaling of the normalisation with $m_{\rm sub}(z=0)$
follows the resolved trend. Recall that these are not the masses at collapse 
so to put the circle and star symbols on the right of Fig.~\ref{Fig_boost} we use $m_{\rm sub}(z=0)\sim10^6$M$_\odot\rightarrow m_{\rm min}\sim10^7$M$_\odot$ 
as explained two paragraphs above, while the location of the star symbol on the left of the figure is somewhat uncertain.
The triangle symbol is the extrapolation to lower masses made by \citet{Kamionkowski_10} using the probability 
distribution function of the density field (subhaloes imprint a power-law tail in this PDF) calibrated with the MW-sized simulation Via Lactea II 
\citep{Diemand_08}. The results from our 
method are quite close to those found by \citet{Kamionkowski_10} and are an order of magnitude lower than the estimates by \citet{Springel_08}.

\citet{Zavala_10} also estimated a subhalo boost with a statistical analysis of all haloes in the Millennium-II simulation 
\citep{BK_09}, implicitly extrapolating the subhalo mass function and concentration-mass relation to the unresolved regime. 
They found a large range of possible subhalo boosts, within $2-2\times 10^3$, depending on the exact parameters of the extrapolation,
for a $\sim10^{12}$M$_{\odot}$ halo. 

\subsection{Sommerfeld-enhanced models}\label{SE_models}

If the annihilation cross-section is enhanced by a Sommerfeld mechanism \citep[e.g.][]{Arkani_09,Pospelov_09}, then the annihilation rate increases with 
lower relative velocities until saturating due to the finite range of the interaction between the particles prior to annihilation. Instead of using a 
specific Sommerfeld-enhanced model, we generically approximate the cross section as:
\begin{eqnarray}\label{SE_eq_0}
  (\sigma_{\rm ann} v)&=&(\Delta v/c)^{-\beta_{\rm S}}(\sigma_{\rm ann} v)_0, \ \ \ \ \ \Delta v>\Delta v_{\rm sat}\nonumber\\
  (\sigma_{\rm ann} v)&=&S_{\rm sat}(\sigma_{\rm ann} v)_0, \ \ \ \ \ \Delta v\leq\Delta v_{\rm sat}
\end{eqnarray}
The value of $\beta_{\rm S}$ is commonly near $1$ (the so-called ``$1/v$'' boost), but it can reach $2$ near resonances; we only consider the former case.
For Eq.~(\ref{SE_eq_0}) and using our modelling we obtain:

\begin{widetext}
\begin{eqnarray}\label{SE_eq}
   R_{\rm ann}^{\rm subs}(SE)&=&\frac{8\pi^{1/2}b^3}{9\delta_c^3}200\rho_{crit,0}M_{200}\frac{(\sigma_{\rm ann} v)_0}{2m_\chi^2}
   \times \Bigg[ S_{\rm sat}\int_{\rm m_{min}}^{\rm m_{sat}} \mu(m_{\rm col})m_{\rm col}^{-2}d(m_{\rm col}^2\sigma^3(m_{\rm col}))\nonumber\\
     &+& \left(\frac{\left(10H_0G\right)^{2/3}}{\pi\delta_c c^2}\right)^{-\beta_{\rm S}/2}
     \int_{\rm m_{sat}}^{\rm m_{max}} \left(m_{\rm col}^{2/3}\sigma(m_{\rm col})\right)^{-\beta_{\rm S}/2}
    \mu(m_{\rm col})m_{\rm col}^{-2}d(m_{\rm col}^2\sigma^3(m_{\rm col}))\Bigg]
\end{eqnarray}
\end{widetext}
\noindent where $m_{\rm sat}$ is the collapse mass corresponding to the characteristic velocity at saturation $\Delta v_{\rm sat}$; using Eq.~(\ref{lame_mod}):
\begin{equation}
  \Delta v_{\rm sat} = b\left(10H_0Gm_{\rm sat}\right)^{1/3}\left(\frac{\sigma(m_{\rm sat})}{\pi\delta_c}\right)^{1/2}.
\end{equation}

\begin{figure}
\center{
\includegraphics[height=8.0cm,width=8.0cm]{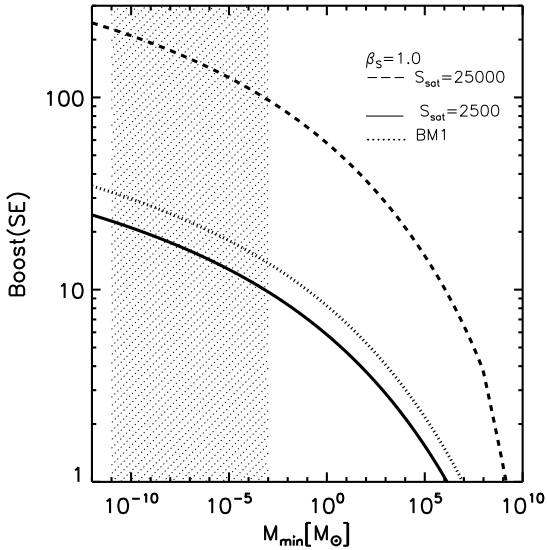}
}
\caption{Same as Fig.~\ref{Fig_boost} but for a Sommerfeld-like model where $(\sigma_{\rm ann} v)=(\Delta v/c)^{-\beta_{\rm S}}(\sigma_{\rm ann} v)_0$ with $\beta_{\rm S}=1.0$ and different
saturation values as shown in the legend. The dotted line is an approximation to one of the benchmark points (BM1) presented in \citet{Finkbeiner_2011} 
that avoid different astrophysical constraints and can account for high energy cosmic ray excesses.The dotted region shows the theoretical expectation 
for the decoupling mass limit of WIMPs \citep{Bringmann_09}.} 
\label{Fig_boost_SE} 
\end{figure}
Fig.~\ref{Fig_boost_SE} shows the subhalo boost, relative to the smooth dark matter distribution of the host halo, for Sommerfeld-like models where
the cross section scales as in Eq.~(\ref{SE_eq_0}) with $\beta_{\rm S}=1$. To compute the Sommerfeld-enhanced smooth component, we simply took the annihilation 
rate corresponding to the Einasto profile, defined in Section \ref{sec_rate}, and scaled it up based on the characteristic velocity dispersion
of the Aq-A-2 halo: $R_{\rm ann}^{\rm smooth}(SE)=S(\sigma_{\rm disp}^{\rm host})R_{\rm ann}^{\rm smooth}$, with 
$\sigma_{\rm disp}^{\rm host}\sim120{\rm km s}^{-1}$. The solid line is for $S_{\rm sat}=2500\sim S(\sigma_{\rm disp}^{\rm host})$; below this value, all substructures are 
saturated and thus the boost is just a trivial scaled version of that shown in Fig.~\ref{Fig_boost}. The dotted line roughly approximates one of
the benchmark point models defined in \citet{Finkbeiner_2011} (BM1 in their Table 1) to satisfy a number of astrophysical constraints (relic abundance, CMB
power spectrum, etc.) while at the same time providing a fit to the cosmic ray excesses observed by the PAMELA and Fermi satellites. Enhancements 
much larger than this value (e.g. dashed line, $S_{\rm sat}=20000$) are likely ruled out by astrophysical constraints but illustrate the 
transition between the unsaturated and saturated regimes. 
We note that BM1 is still compatible with the new measurement of the positron excess reported by the AMS 
collaboration \citep[see][]{Cholis_2013}. Since allowed Sommerfeld-enhanced models in the parameterisation we have used have 
$S_{\rm sat}\gtrsim S_{\rm host}$, then the enhanced substructure boost is at the end just a scaled version of the constant cross section boost: 
${\rm Boost(SE)}\sim S_{\rm sat}/S_{\rm host}\times {\rm Boost (\sigma_{\rm ann} v = const)}$.

\subsection{Halo-centric distance boost}

Finally, we can also apply our methodology to compute the substructure boost as a function of the distance to the halo centre. To do so, all that is needed
is to replace the global values of the parameters $b$ and $\mu(m_{\rm col})$ for the radial dependent values that we fit to the simulation
data in Section \ref{sim_fit}, and to account for the renormalisation in mass of the phase-space volume where \psad~is averaged. A simple but good 
approximation for $\mu(r;m_{\rm col})$ is given in Eq.~(\ref{mu_r}). Using it, our results can be easily 
reproduced with the values given in Table \ref{table_fit_radial}.

\begin{figure}
\center{
\includegraphics[height=8.0cm,width=8.0cm]{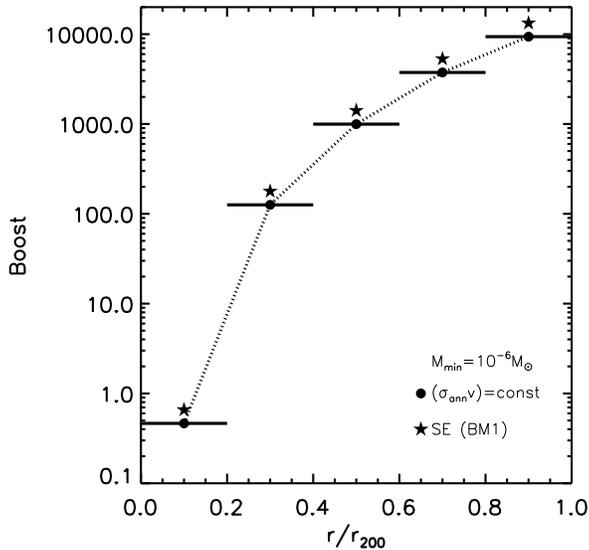}
}
\caption{Subhalo boost to the annihilation rate as a function of halo-centric distance for a minimum subhalo mass {\it at collapse} of $10^{-6}$M$_{\odot}$.
The boost was computed for concentric shells with a thickness of $0.2~R_{200}$ in the cases of constant cross section (circle symbols) and for an approximation
of the Sommerfeld-enhanced model BM1 presented in \citet{Finkbeiner_2011} (star symbols).} 
\label{Fig_boost_radial} 
\end{figure}
Fig.~\ref{Fig_boost_radial} shows the substructure boost to the annihilation rate of the smooth halo as a function of the distance to the halo centre. Since
in Section \ref{hc_dist_sec} we calibrated our fitting parameters in concentric shells with a thickness of $0.2~R_{200}$, we plot the results with bars showing
the extent of the shells in which \psad~was averaged. We remark that in this case $M_{{\cal V}_6}$ is the total mass in a given shell, not $M_{200}$ as it has
been so far in this section. In Fig.~\ref{Fig_boost_radial} we show the cases with $(\sigma_{\rm ann} v)={\rm const}$ (circles) and Sommerfeld-like enhancement corresponding 
to the benchmark point BM1 (stars) presented in \citet{Finkbeiner_2011} (see Section \ref{SE_models}). Our results are in good agreement with those of 
\citet{Kamionkowski_10} (see their Fig. 4), although we seem to predict lower boosts in the central regions. 

Values specific to a certain radius can be 
approximated by interpolating $b(r)$ and $\mu(r;m_{\rm col})$ and considering that for small volumes one has 
$R_{\rm ann}^{\rm smooth}(r)\propto\rho_{\rm smooth}(r)M_{{\cal V}_6}$ and thus the subhalo boost can be computed independently of the specific value of $M_{{\cal V}_6}$.
For instance, for $r=8$~kpc, we find that the boost is only $\sim0.1\%$ for the constant cross section case, and $\sim0.2\%$ for the Sommerfeld-like model
BM1.

\section{Summary and Conclusions}\label{conc_sec}

The clustering of dark matter at scales unresolved by current numerical simulations is a key ingredient in many predictions of non-gravitational (and some gravitational) signatures
of dark matter. The characteristic scale of the smallest haloes (commonly called microhaloes) contributing to these signals is $\mathcal{O}(10^{-9})$ times the size of the Milky Way halo, and we
therefore refer to it as the {\it nanostructure} of dark matter haloes.
The degree of uncertainty of this {\it nanostructure} clustering can be as much as two orders of magnitude for WIMP dark matter models, since the minimum
bound haloes have masses $\gtrsim9$ orders of magnitude below the highest resolution simulations to date. In the case of dark matter annihilation for example,
the small, cold and dense dark structures have the dominant contribution to the hypothetical extragalactic signals, and thus the bulk of the
predicted annihilation rate comes from extrapolating, in diverse ways, the dark matter clustering from the resolved to the unresolved regime.

Most extrapolation methods rely on assumptions about the abundance, spatial distribution, and internal properties of (sub)haloes. For example, the total subhalo 
boost to the annihilation rate over the smooth dark matter distribution of a single halo, is typically computed by extrapolating, either explicitly or implicitly,
the subhalo mass function and the concentration-mass relation. Extending these functions as power-laws down to lower masses yields the higher boosts. In this
paper, we present an alternative method based on a novel perspective on the clustering of dark matter introduced in \citet{Zavala_13} (Paper I). The method rests upon
writing the dark matter annihilation rate in a given volume as an integral over velocities of a new coarse-grained Particle Phase Space Average Density 
(\psad; see Eqs.~\ref{coarse} and \ref{rate}). 

The structure of \psad~was analysed in detail in Paper I, where it was found to be nearly universal
in time and across different ambient densities, in the regime dominated by substructures. Here we present a model of the structure of \psad~inspired by
the stable clustering hypothesis and the spherical collapse model, and improved by incorporating a prescription for the tidal disruption of subhaloes.

Our modelling provides a physically-motivated explanation of the two-dimensional functional shape of \psad, and calibrated to the MW-size Aquarius simulations, gives
a firm basis to extrapolate into the unresolved substructure regime. In summary, the main advantages of our method are two-fold: 
\begin{itemize}
\item The free parameters in our model are fitted to a single 2D function that is a very sensitive measure of cold small scale substructure and is directly
  connected to the annihilation rate. 
\item The annihilation rate is written as an integral over relative velocities of the ``zero-separation'' limit of \psad~multiplied by the annihilation cross section
  times the relative velocity (see Eq.~\ref{rate}). This allows us to accommodate any velocity-dependence coming from particle physics models in a straightforward way.
\end{itemize}
Although our model has several free functions that are fitted to the MW-size Aquarius simulations, 
only two are of relevance for the prediction of the annihilation rate: (i) $b$ in Eq.~(\ref{lame_mod})
parametrizes the ratio of phase space volume in spherical collapse, to the characteristic phase space volume of subhaloes at 
the time of formation; and (ii) $\mu$ in Eq.~(\ref{ellipse_0}) that can be interpreted as the dilution of the characteristic phase space density at
collapse due to tidal disruption. For the simulated data we analised, the former is $\sim1.8$ and we present tabulated fitted 
values across redshifts and distances to the MW-halo centre (Tables 2 and 3), while for the latter we present simple parameterisations (see Tables 2 and 3,
Eqs.~29-30).

As a sample application of our model, we computed the subhalo boost (over the smooth dark matter distribution) in a MW-size halo, both globally 
(i.e. over the whole virialized halo) and locally as a function of halo-centric distance, for cases where the annihilation cross section is constant and 
for a generic Sommerfeld-like enhanced case where $(\sigma_{\rm ann} v)\propto1/\Delta v$ up to a maximum saturation value. We find that in the former, the global boost
is $\sim15$ for typical WIMP models (with decoupling masses in the range $10^{-11}-10^{-3}$M$_{\odot}$). 

Our estimate of the subhalo boost is in the low end of current estimates
being in agreement with \citet{Kamionkowski_10} (based on a different simulation), and a factor of $\sim10$ lower than the boost computed in \citet{Springel_08}
(based on the same simulation suite than the one used here). The discrepancy with the latter is likely caused by their implicit extrapolation of the subhalo mass
function and, perhaps more importantly, the concentration-mass relation. Evidence of this can be seen in the structure of \psad~in the resolved 
regime (without recurring to the modelling of the unresolved regime) where our analysis indicates that the annihilation rate 
(in the $(\sigma_{\rm ann} v)={\rm const}$ case) from 
substructures scales logarithmically with the maximum (coarse-grained) phase space density (Eq.~\ref{rate_cte}) rather than as a power law. 
Since we argue that \psad~is a more direct measure of the annihilation signal, our results seem to disfavour large boost factors.

\psad~has also other potential applications in dark matter direct detection, pulsar timing and transient weak lensing \citep[see][]{Rahvar_13} that will be
explored in the future. 

We have made our code to compute \psad~(with our full model) publicly available online at {\rm http://spaces.perimeterinstitute.ca/p2sad/}. Interested users should refer to 
that site for instructions on how to use the code. 

\section*{Acknowledgments}

We thank the members of the Virgo consortium for access to the Aquarius simulation suite and our special thanks to
Volker Springel for providing access to SUBFIND and GADGET-3 whose routines on the computation of the 2PCF in
real space were used as a basis for the 2PCF code in phase space we developed.
JZ and NA are supported by the University of 
Waterloo and the Perimeter Institute for Theoretical Physics. Research at Perimeter Institute is supported by the Government of Canada 
through Industry Canada and by the Province of Ontario through the Ministry of Research \& Innovation. JZ acknowledges financial support by a 
CITA National Fellowship. This work was made possible by the facilities of the Shared Hierarchical 
Academic Research Computing Network (SHARCNET:www.sharcnet.ca) and Compute/Calcul Canada. The Dark Cosmology
Centre is funded by the DNRF.

\bibliography{lit}

\end{document}